# Biaxial strain effect of spin dependent tunneling in MgO magnetic tunnel junctions

Ajeesh M. Sahadevan,[1] Ravi K. Tiwari,[2] Gopinadhan Kalon,[1] Charanjit S. Bhatia,[1] Mark Saeys,[2,a)] and Hyunsoo Yang[1,a)]

[1]Department of Electrical and Computer Engineering, National University of Singapore, 4 Engineering Drive 3, Singapore 117576, Singapore

[2]Department of Chemical and Biomolecular Engineering, National University of Singapore, 4 Engineering Drive 4, Singapore 117576, Singapore

We study the effect of strain on magnetic tunnel junctions (MTJ) induced by a diamond like carbon (DLC) film. The junction resistance as well as the tunnel magnetoresistance (TMR) reduces with the DLC film. Non-equilibrium Green's function quantum transport calculations show that the application of biaxial strain increases the conductance for both the parallel and anti-parallel configurations. However, the conductance for the minority channel and for the anti-parallel configuration is significantly more sensitive to strain, which drastically increases transmission through a MgO tunnel barrier, therefore, the TMR ratio decreases with biaxial strain.

[a)] e-mail address: chesm@nus.edu.sg and eleyang@nus.edu.sg

The material and interface engineering of magnetic tunnel junctions (MTJ) is the key for the future of spin transfer torque based random access memories (STT-RAM). A value of tunneling magnetoresistance (TMR) exceeding 1000% has been predicted[1,2] and recently achieved in MTJs with single crystalline or textured MgO(001) tunnel barriers[3-5] through which the $\Delta_1$ Bloch states tunnel coherently. In this case, the crystalline property of the structure, especially that of the ferromagnet/MgO interface plays a very critical role in the device performance[6,7] and the presence of strain in the structure can change the properties of the device significantly. Strain has been used to improve the performance of semiconductor devices such as the metal oxide semiconductor field effect transistors (MOSFETs) without failure at low cost over the last decade.[8] Recently, a line stressor based on diamond like carbon (DLC) films has been proposed with very high intrinsic stress (few GPa) and high $sp^3$ content.[9,10]

For the case of MgO based MTJs, the role of epitaxial strain has been discussed previously.[11,12] Yeo *et al.* studied the interface states of a strained MgO/Fe(001) system and showed the position of the minority spin peak state near the Fermi energy shifts upwards in energy with respect to the Fermi energy for tensile strain, while it moves downwards for compressive strain.[12] A few experimental reports have been also studied the effect of lattice mismatch on the transport properties in Fe/MgO/Fe(001) and related systems.[13-15]

In this work, we study the effect of a DLC film on the tunneling behavior of MgO based MTJs. With the deposition of DLC film over tunnel junctions, the TMR as well as the junction resistance is suppressed showing the effect of external mechanical strain on the transport properties. Though the TMR is reduced, it is interesting to understand the physics behind it and the strain induced reduction of the junction resistance is encouraging for industries utilizing MTJs as read sensors in hard disk drives. To corroborate the experimental results, the effect of biaxial strain is evaluated using Non-equilibrium Green's function (NEGF) quantum transport calculations. The minority and the anti-parallel transmission both increase more than the majority transmission, and biaxial compressive strain is calculated to decrease the TMR ratio, in agreement with the experiments.

MTJs have been grown using magnetron sputtering in an ultra-high vacuum chamber with the structure of 100 Ta/300 $Ir_{22}Mn_{78}$/6 $Co_{40}Fe_{40}B_{20}$/30 $Co_{70}Fe_{30}$/8 Ru/27 $Co_{70}Fe_{30}$/8 Mg/14 MgO/20 $Co_{40}Fe_{40}B_{20}$/50 Ta/50 Ru (all thickness in Å). The MgO barrier is formed by the reactive sputter deposition of Mg in Ar-$O_2$ plasma (~2% oxygen). Samples are annealed at 300 °C for 30 minutes under 1 T magnetic field, and then MTJs are fabricated in a current perpendicular-to-plane (CPP) configuration using a combination of Ar ion-milling and photolithography processes. A number of devices of different junction areas are measured after fabrication and a 40 nm DLC film is then deposited over the junction. Figure 1(a) shows a schematic diagram of the device with the DLC layer on top, exerting a compressive biaxial strain on the tunnel junction along *x* and *z* axes[9], while Fig. 1(b) shows a scanning electron microscope (SEM) image with a DLC film over a tunnel junction. DLC films are grown by filtered cathodic vacuum arc and a method used by Ehsan *et al.* has been adopted that provides good adhesion along with high $sp^3$ content for enough strain.[16] Ion energy of about 100 eV is selected as it provides the highest fraction of $sp^3$ bonds with the maximum density and hardness.

Figure 1(c) shows the X-ray photoelectron spectroscopy (XPS) spectra of the $C_{1s}$ core level for the DLC film, which indicates a very high $sp^3$ proportion (65%) of the film as has been used for MOSFETs with compressive stress as high as 7.5 GPa.[9] A higher $sp^3$ fraction in the DLC film is important to induce enough strain. The atomic fraction of each component (C-C $sp^1$,

C-C $sp^2$, C-C $sp^3$, C=O, and C-O) is obtained by integrating the associated Gaussian curves deconvoluted from the $C_{1s}$ spectra after Shirley background subtraction.[16]

Figure 1(d) shows a plot of the TMR versus the junction area before and after the deposition of DLC film at room temperature. The TMR ratio is defined by $(R_{AP}-R_P)/R_P$, where $R_P$ and $R_{AP}$ are the junction resistance in the parallel (P) and anti-parallel (AP) alignment of the ferromagnetic electrodes, respectively. Before the deposition of the DLC film, it is observed that the TMR gradually reduces as the junction area increases. The DLC film is then deposited over tunnel junctions of areas ranging from 50 to $10^4$ µm$^2$. It is clearly observed that there is a suppression of TMR for junction areas below 500 µm$^2$, and that the TMR after the DLC deposition gradually reduces as the junction area decreases. The change due to the DLC layer is bigger in the devices with smaller junction areas due to higher effective strain in a smaller junction. The free layer loop of the device before and after the DLC deposition has no difference, therefore we rule out any possibility of pinholes in our samples due to strain, otherwise a shift in the free layer loop at lower temperatures is expected [17]. We have also done transport of ions in matter calculations for the energy of C ions used in this study. The maximum penetration depth of C in the Cu top electrode (100 nm thick) is only 5 nm with a peak at ~ 1 nm, which also support that the junction damage due to the DLC deposition is negligible.

The voltage dependence of the TMR and junction resistance show a tunneling feature before the DLC deposition as shown in Fig. 2(a), such that the TMR decreases with increasing bias voltage and $R_P$ is independent of bias voltage, as typically observed in MgO based MTJs.[18] Figure 2(b) shows suppression in the voltage dependence of TMR after the deposition of the DLC film over the device. For the unstrained device, the relative reduction in TMR at 0.4 V is 43% with respect to the value at zero bias, while for the strained device the relative reduction is only 14%. We have also carried out temperature dependent studies. For the unstrained device, $R_P$ has little temperature dependence, while $R_{AP}$ increases as the temperature reduces shown in Fig. 2(c). On the other hand, when the same device is strained using the DLC film, in addition to reduction in the magnitude of TMR and $R_{AP}$, their temperature dependence is also suppressed as shown in Fig. 2(d). These observations can be related to the changes in the tunneling probabilities as shown by our calculations later.

Coherent tunneling transport in a Fe/MgO/Fe tunneling junction [Fig. 3(c)] is described by the NEGF formalism[19] as implemented in the Green program.[20, 21] The electronic structure of the tunneling junction is described by an Extended Hückel Molecular Orbital (EHMO) Hamiltonian[22], using literature values for the Fe *spd*, Mg *spd*, and O *sp* parameters.[23] With those parameters, a bulk MgO band gap of 7.8 eV and a bulk Fe magnetic moment of 2.0 $\mu_B$ are calculated, in good agreement with experimental data of 7.77 eV and 2.2 $\mu_B$, respectively,[24, 25] and with hybrid density functional theory calculations using the HSE03 functional (DFT-HSE03)[26-28]. Detailed DFT-HSE03 calculations for a 4-layer MgO slab on a 6-layer Fe(100) slab show that the top of the MgO valence band edge is located about 4.0 eV below the Fermi level of the system. Therefore, a similar offset is used in our EHMO-based transport calculations. Note that the EHMO band gap for a 6-layer MgO slab, 7.2 eV, is significantly larger than the DFT-HSE03 value of 3.7 eV. The effect of biaxial strain on the MgO band gap is, however, accurately described by EHMO, as discussed below.

Transport calculations are performed for MgO barriers of 4, 6, 8, 10, and 12 layers. For the unstrained transport calculations, the experimental Fe lattice constant of 2.87 Å is used for the Fe(100) contacts. MgO(001) slab, rotated 45 degrees with respect to the Fe lattice, is then placed 2.16 Å over Fe (100) contacts so that O atoms sit directly above Fe atoms. Assuming a biaxial stress in the range of 5 to 10 GPa leads to a compression of about 2.5 to 5%. Therefore, the effects of compression on the conductance and the TMR ratio are evaluated for a compression of 5% along the *x* and *z* direction [as defined in Figure 3(c)], and an expansion of 2.3% in the y direction using the MgO Poisson ratio of 0.187.[29, 30] In reality, the compression is likely somewhat smaller than 5%. To confirm that our results remain valid for different amounts of strain, transport calculations are also performed for a 6-layer MgO junction with 2.5%, 3.5% and 10% compressive strain.

The effect of 5% biaxial *xz*-strain on the junction conductance is calculated as shown in Fig. 3(a) for different MgO barrier thicknesses. For the unstrained junction, both the P and AP conductance decrease exponentially with the number of MgO layers, as expected for tunneling transport. Though the conductance is sensitive to the details of the calculations, our decay rate of 0.40 Å$^{-1}$ for the P configuration and 0.50 Å$^{-1}$ for the AP configuration agree well with published values of 0.44 and 0.49 Å$^{-1}$, respectively[23]. The different decay rate can be understood from the $k_{//}$-resolved transmission spectra in Fig. 4. While majority-to-majority transport is dominated by states near the gamma point, no such states are available for minority-to-minority transport in agreement with earlier calculations.[2] This can be understood from the Fe(100) surface spectral density at the Fermi energy [Fig. 5] and from the complex band structure of MgO.[2] The complex MgO band structure shows that the decay rate is minimum in a small region around the gamma point and increases away from the gamma point[2]. Therefore, the decay rate in the MgO junction is higher for the minority states.

The application of 5% biaxial *xz*-strain increases the conductance for both the P and AP configuration [Fig. 3(a)]. However, the increase is more pronounced for the AP configuration, and hence the TMR ratio decreases by a factor 10 to 30 [Fig. 3(b)]. Biaxial strain decreases the decay constants to 0.37 and 0.45 Å$^{-1}$ for the P and the AP configuration, respectively. In addition, the contact conductance, a measure of the number of active transport channels in the Fe(100) contacts and their coupling at the Fe(100)/MgO interface[31], increases 10-fold for the minority channels, whereas only 2-fold for the majority channels. The increase in the conductance by a factor 1.3 to 3.7 and the decrease in decay rate for the P configuration can be attributed to a decrease in the MgO band gap. Indeed, the EHMO band gap for a 6-layer MgO slab decreases from 7.29 eV to 7.02 eV, comparable to the 0.11 eV decrease calculated by DFT-HSE03. The conductance for the AP configuration is more sensitive to biaxial strain, and increases by a factor 7 for a 4-layer MgO barrier and 5% strain, and by a factor 61 for a 12-layer MgO barrier.

To compare the calculated changes in the TMR ratio and in the conductance with the experimental values, we have included the relative changes in Fig. 3. The 22-fold increase in the AP conductance for 5% strain and for a 6-layer (13 Å) MgO barrier is significantly larger than the experimental increase of 2.9 for a 20 Å MgO barrier. Also the 15-fold decrease in the calculated TMR ratio for a 6-layer MgO barrier is larger than the experimental value of 4.8 for a 20 Å MgO barrier. However, when the biaxial strain is reduced to 3.5% for the 6-layer MgO junction, the agreement with the experiments improves. The calculated increases in the P and in the AP conductance by factors 1.1 and 3.0, respectively, can be compared with the experimentally measured increases by factors 1.7 and 2.9, respectively [Fig. 2]. Also the 2.7-fold decrease in the calculated TMR ratio matches the experimental value of 4.8 quite well. The qualitative implications are preserved even for smaller levels of biaxial strain in the MgO layer.

The more pronounced increase in the conductance for the minority channels and for the AP configuration can be understood from the $k_{//}$-resolved transmission spectra, shown in Fig. 4 for a 6-layer MgO junction in the case of unstrained and a 3.5% biaxial strain. For the unstrained junction, majority transport is dominated by states at the gamma point. The $k_{//}$-resolved spectra for the majority states are relatively unaffected by strain, except for a small broadening of the peak and an increase in the peak maximum from $0.846 \times 10^{-3}$ to $0.853 \times 10^{-3}$. Minority transport, however, is dominated by a circle of states around the gamma point and by states near the Brillouin zone edge. Biaxial *xz*-strain breaks the 4-fold symmetry in the *xy* plane, and transmission hot-spots move closer to the $k_y = 0$ axis. The change in the location of minority states at the Fermi energy is also illustrated by the Fe(100) spectral density in Fig. 5. The minority states are concentrated in a narrow square region around the gamma point, with few states at the gamma point. In this region the decay constant for MgO is quite high. Biaxial strain increases the orbital overlap in the *x* direction and hence broadens the *d*-band. As can be seen in Fig. 5, this moves minority states closer to the gamma point along the $k_x$ axis. The effect is even more pronounced in the transmission spectrum [Fig. 4] and new hot-spots appear near the $k_y=0$ axis. The change in the minority states is also reflected in a decrease in the Fe magnetic moment. For a 6-layer Fe(100) slab, the EHMO magnetic moment per Fe decreases from 1.96 to 1.86 $\mu_B$, again in good agreement with the 0.14 $\mu_B$ decrease calculated by DFT-HSE03.

The above results also help to explain the experimental voltage and temperature dependence of the TMR. It was reported that the transmission for the minority spin channel is sensitive at low biases, however once the minority states moves closer to the gamma point, the bias dependent transmission is significantly suppressed at higher biases.[32] From our calculations we conclude that the biaxial strain moves the minority states closer to the gamma point. This in turn weakens the sensitivity of the minority states to the voltage and temperature, resulting in a diminished voltage and temperature dependence of $R_{AP}$ and TMR.

We have demonstrated the effect of mechanical stress on the tunnelling properties of MgO tunnel junctions. The deposition of a DLC film with a very high intrinsic stress over the junction reduces the TMR ratio as well as the junction resistance. The NEGF calculations reproduce both the increase in the conductances and the decrease in the TMR ratio, when biaxial *xz*-strain is applied. The increase in the conductance for the parallel configuration can be attributed to a decrease in the MgO band gap by about 0.3 eV and the barrier thickness by 5%. The conductance for the anti-parallel configuration is significantly more sensitive to strain, which is attributed to changes in the location of the Fe(100) minority states at the Femi level. When strain is applied, the *d*-band broadens and the minority states at the Fermi energy move closer to the center of the 2D Brillouin zone where transmission through the MgO barrier is higher. As a result, hot-spots appear in the $k_{//}$-resolved transmission spectrum and the conductance for both the minority channel and for the anti-parallel configuration increase rapidly. This study demonstrates the important effect of strain on the anti-parallel conductance, and suggests that strain can reduce the resistance-area product for MgO based read sensors with a sufficiently high TMR value.

This work was supported by the Singapore NRF CRP Award No. NRF-CRP 4-2008-06.

**Figure captions**

FIG. 1. (a) Schematic of the device with a DLC layer over the junction. (b) An SEM image with a DLC film. The top electrode width is 80 μm while the DLC strip has a width of 150 μm. (c) XPS spectra of the $C_{1s}$ core level for the DLC film. (d) A plot of TMR versus junction area.

FIG. 2. Bias voltage dependence of $R_P$, $R_{AP}$, and TMR for MTJ before (a) and after (b) the DLC deposition at 300 K. Temperature dependence of $R_P$, $R_{AP}$, and TMR before (c) and after (d) the DLC deposition, for a device with the junction area of 73 μm$^2$.

FIG. 3. (a) Calculated conductance for a Fe(100)/MgO/Fe(100) tunneling junction as a function of the number of MgO layers. The conductance is shown for the P and the AP configurations for both the unstrained and for 5% biaxial *xz*-strain cases. The relative increase in the conductance after applying strain is also shown to facilitate comparison with the experimental data in Fig. 2. For 6 MgO layers, the P conductance increases by a factor 1.74 from 0.65 to 1.14 nS, while the AP conductance increases by a factor 22.32 from 7 to 157 pS. (b) Optimistic TMR ratio [($G_P$-$G_{AP}$)/$G_{AP}$, where $G_P$ and $G_{AP}$ is the conductance of the P and the AP state, respectively] for the unstrained and the strained tunneling junction. The relative change in the TMR ratio is also shown and ranges from a factor 7 to 27. (c) Central structure used to model the junction for 6 layers of MgO. The blue, green, and red circles correspond to Fe, Mg, and O atoms, respectively. In the calculations, both Fe(100) contacts extend to infinity.

FIG. 4. $k_{//}$-resolved transmission spectra for the various transport modes for a Fe(100)/MgO(6 layers)/Fe(100) junction. Biaxial strain decreases the lattice in the *x* and *z* direction by 3.5%, and expands the lattice in *y* direction by 1.6%. Note the different scales for the various transmission spectra.

FIG. 5. Effect of 3.5% biaxial *xz*-strain on the Fe(100) surface spectral density (number of states/eV/Å$^2$) at the Fermi energy for the minority and the majority states. While changes for the majority states are relatively minor, the minority states at $(k_x, k_y)=(\pm 0.4, 0.0)$ clearly move closer to the gamma point. This is consistent with a broadening of the minority band and a decrease in the spin polarization.

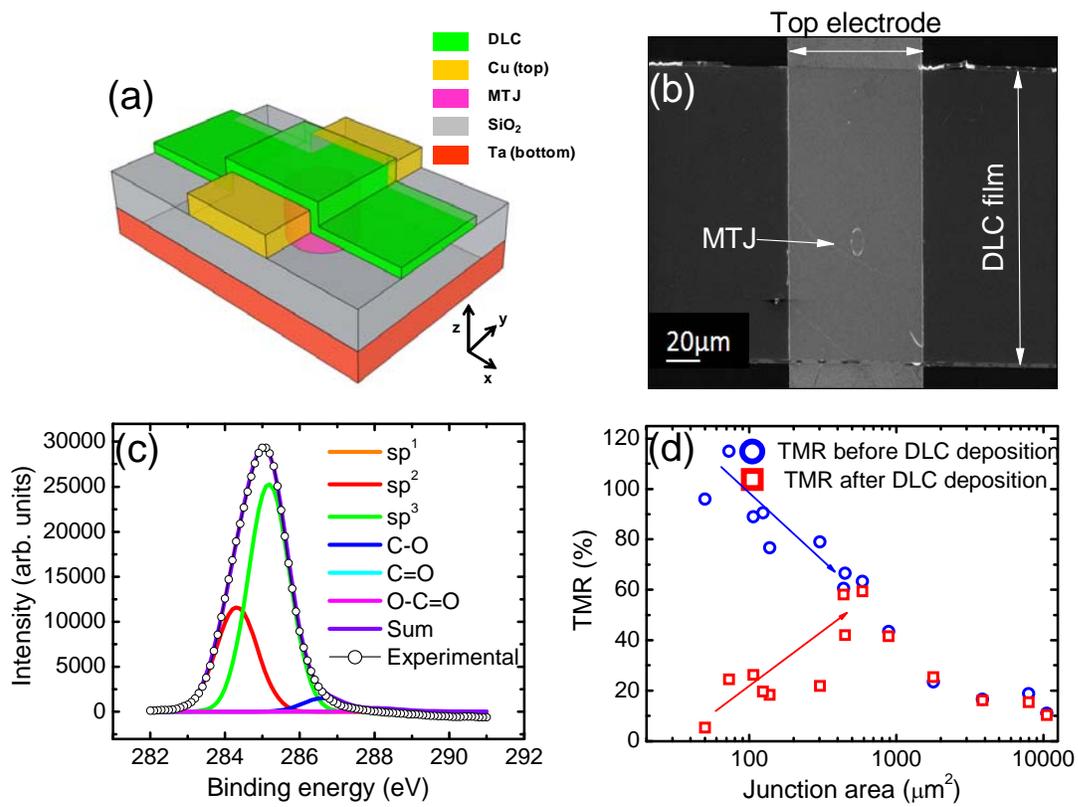

Figure 1

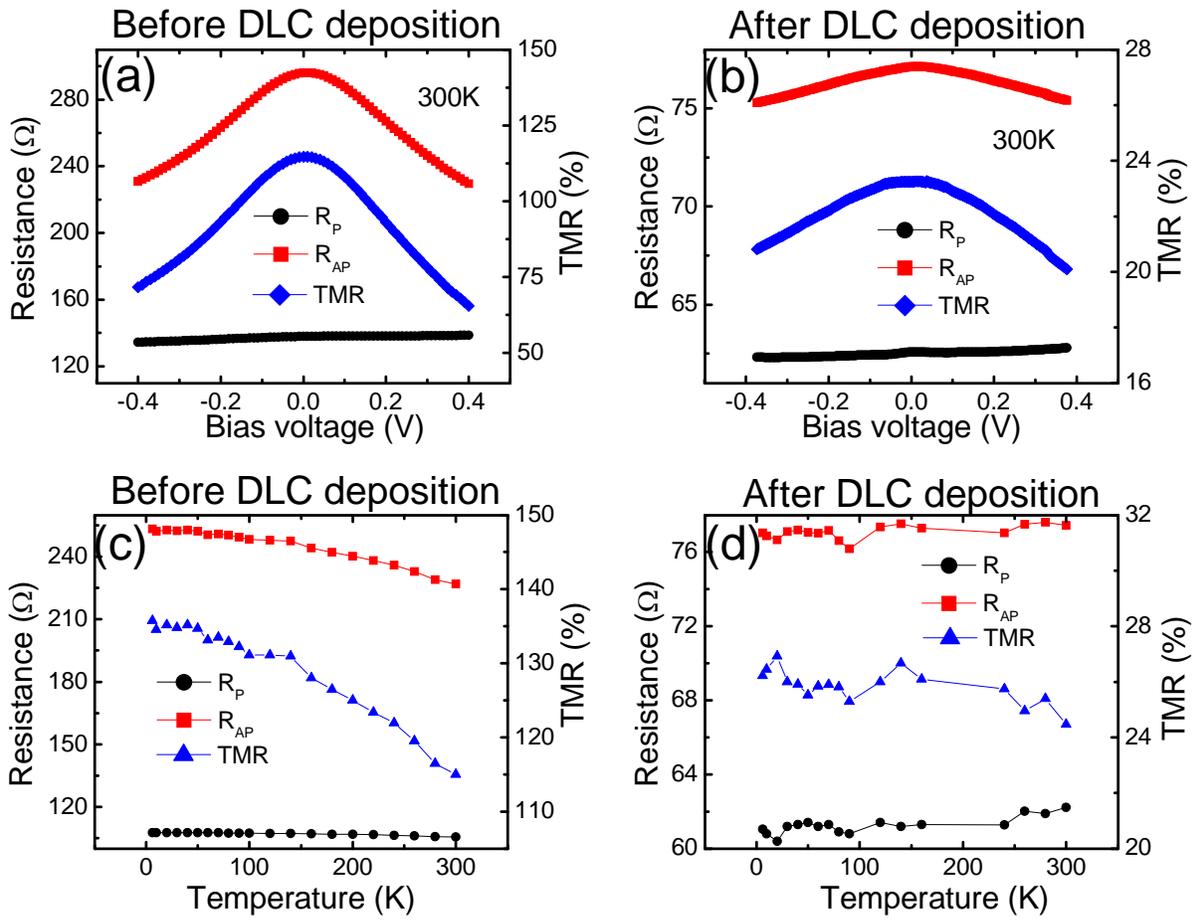

Figure 2

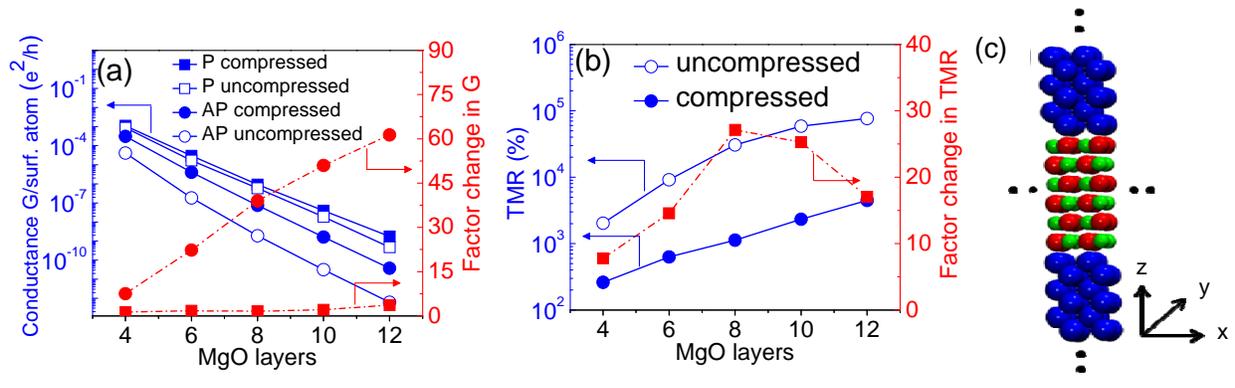

Figure 3

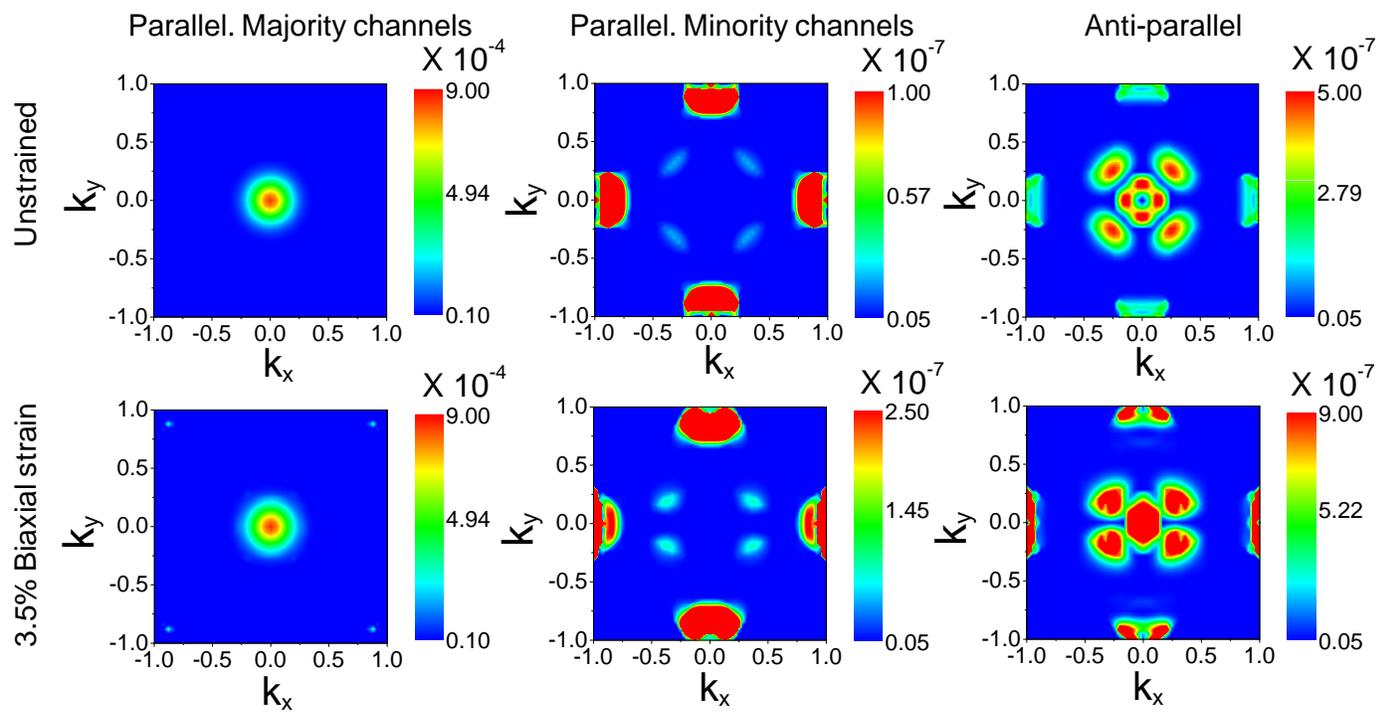

Figure 4

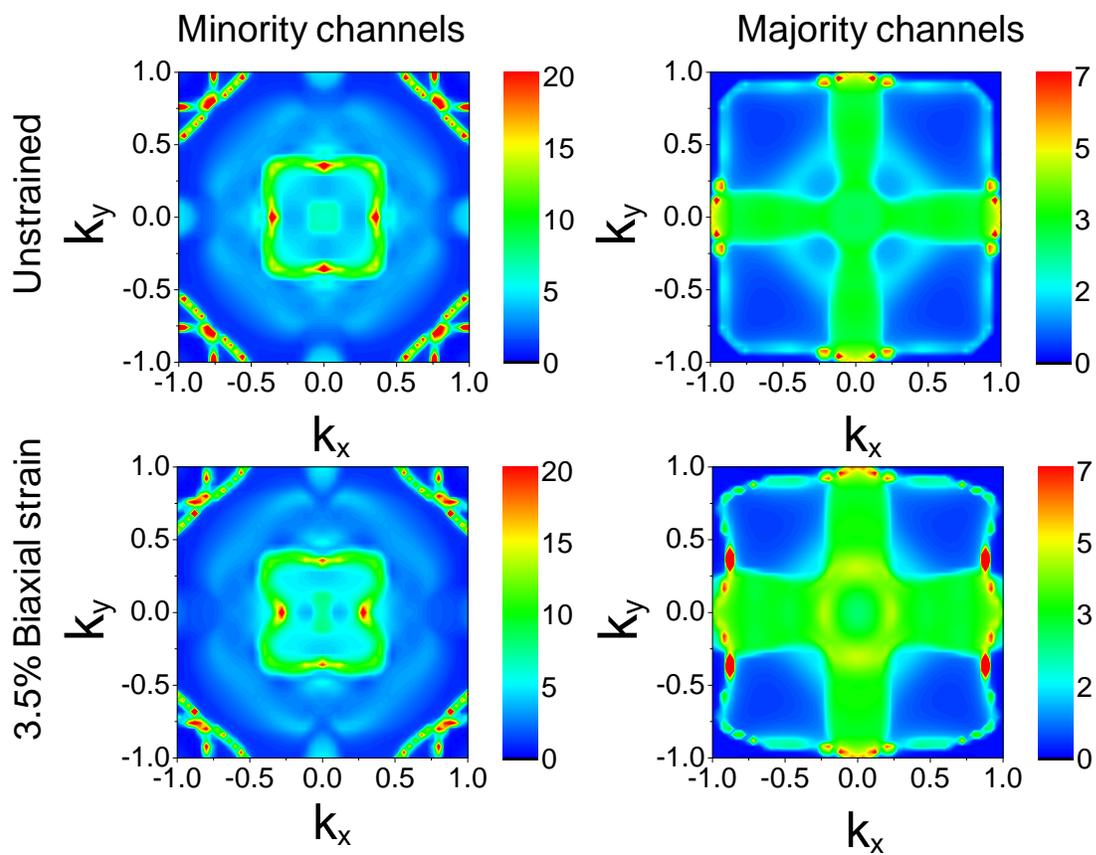

Figure 5